# Self-Estimation of Path-Loss Exponent in Wireless Networks and Applications

Yongchang Hu, *Member, IEEE*, and Geert Leus, *Fellow, IEEE*

*Abstract*—The path-loss exponent (PLE) is one of the most crucial parameters in wireless communications to characterize the propagation of fading channels. It is currently adopted for many different kinds of wireless network problems such as power consumption issues, modeling the communication environment, and received signal strength (RSS)-based localization. PLE estimation is thus of great use to assist in wireless networking. However, a majority of methods to estimate the PLE require either some particular information of the wireless network, which might be unknown, or some external auxiliary devices, such as anchor nodes or the Global Positioning System. Moreover, this external information might sometimes be unreliable, spoofed, or difficult to obtain. Therefore, a self-estimator for the PLE, which is able to work independently, becomes an urgent demand to robustly and securely get a grip on the PLE for various wireless network applications. This paper is the first to introduce two methods that can solely and locally estimate the PLE. To start, a new linear regression model for the PLE is presented. Based on this model, a closed-form total least squares (TLS) method to estimate the PLE is first proposed, in which, with no other assistance or external information, each node can estimate the PLE merely by collecting RSSs. Second, to suppress the estimation errors, a closed-form weighted TLS method is further developed, having a better performance. Due to their simplicity and independence of any auxiliary system, our two proposed methods can be easily incorporated into any kind of wireless communication stack. Simulation results show that our estimators are reliable, even in harsh environments, where the PLE is high. Many potential applications are also explicitly illustrated in this paper, such as secure RSS-based localization, $k$th nearest neighbor routing, etc. Those applications detail the significance of self-estimation of the PLE.

## I. INTRODUCTION

IN WIRELESS communications, the received instantaneous signal power at receivers is commonly modeled as the product of large-scale path-loss and small-scale fading. Large-scale path-loss fading assumes that the attenuation of the average received power is subject to the transmitter–receiver distance $r$ as $r^\gamma$, where $\gamma$ is the path-loss exponent (PLE). Due to the

Manuscript received March 21, 2014; revised August 8, 2014; accepted December 5, 2014. Date of publication December 10, 2014; date of current version November 10, 2015. This work was supported by the China Scholarship Council (CSC) and Circuits and Systems (CAS) Group, Delft University of Technology, Delft, The Netherlands. The review of this paper was coordinated by Prof. T. Kuerner.
The authors are with the Faculty of Electrical Engineering, Mathematics, and Computer Science, Delft University of Technology, 2628 CD Delft, The Netherlands (e-mail: Y.Hu-1@tudelft.nl; G.J.T.Leus@tudelft.nl).
Color versions of one or more of the figures in this paper are available online at http://ieeexplore.ieee.org.
Digital Object Identifier 10.1109/TVT.2014.2380823

dynamics of the communication channel, the PLE varies over different scenarios and different locations. At the same time, small-scale fading constitutes a rapid fluctuation around the average of the received power and follows a stochastic process. It is mainly due to the multipath effect and changes over very small distances and very small time intervals. However, it can generally be well suppressed by means of some special receiver designs and digital signal processing. Therefore, the PLE becomes a key parameter in characterizing the propagation channel, which significantly determines power consumption, quality of a transmission link, efficiency of packet delivery, etc.

It is important to accurately estimate the PLE so that the wireless communication stack can be dynamically adapted to the PLE changes to yield a better performance. For instance, a path with a relatively low PLE can be chosen to route messages to save power. The PLE is also significant for some other applications. For instance, to calculate the location of a target node in received signal strength (RSS)-based localization, accurate PLE estimation is required, which is mostly provided by reference nodes with known positions. However, in some cases, the reference nodes might be broken and cannot talk to the target node or the location information of the reference nodes might be unreliable, or spoofed by an adversary. Then, accurately estimating the PLE will become a difficult task.

Current methods to estimate the PLE either require some information of the wireless network, which is unknown in most cases, or assistance from auxiliary systems. Three algorithms are presented in [1]: First, when network density is known, the PLE can be estimated by observing RSSs during several time slots and by calculating the mean interference; with regard to the other two algorithms, by changing the receiver's sensitivity, the PLE can be estimated either from the corresponding virtual outage probabilities or from the corresponding neighborhood sizes. All three algorithms require knowledge of the network density or the receiver settings, and even require changing them. Other methods to estimate the PLE mostly lie in the area of RSS-based localization. As already mentioned, using the RSSs for localization requires an accurate estimate of the PLE, which is tightly related with the transmitter–receiver distance. Therefore, special reference nodes with known positions, namely, anchor nodes, are strategically predeployed with the purpose of calibrating the PLE [2]. Considering that the transmitter–receiver distances between anchor nodes can be difficult or expensive to accurately measure in some environments, the PLE can also be estimated by using received power measurements and geometric constraints of anchor nodes to avoid the distance calculation [3]. In the meantime, much effort has been put to jointly estimate the location and the PLE [4]–[6]. Some other methods start with an initial guess of the PLE to approximate the location, which is then used to update the PLE estimate [7], [8]. However, all those methods basically rely on the information from anchor nodes or other auxiliary systems. Once such systems are attacked, unavailable, or generate large errors, the impact on the whole system will be unimaginable. Furthermore, the given methods are also not feasible for many kinds of wireless networks, in which communications and information exchanges might be highly restricted. Therefore, a new self-estimator of the PLE is urgently required, which can solely and locally estimate the PLE without relying on any external assistance. Such an estimator should be able to not only serve localization techniques but act as a general method that can be easily incorporated into any kind of wireless network and any layer of the communication stack as well.

The rest of this paper is structured as follows. In Section II, we present the system model considered in this paper and discuss the problem statement. Some new parameters are introduced in Section III to build a linear regression model for the PLE. Section IV presents and discusses the derivation of our two proposed PLE estimators. Simulation results are given and analyzed in Section V. Many potential applications are discussed in Section VI. Section VII finally summarizes this paper.

## II. SYSTEM MODEL

Here, we introduce some important system model concepts and additionally describe the problem statement.

### A. Node Distribution

Due to the unknown topology of wireless networks, particularly in wireless ad hoc networks, neighbors of a node are ideally considered randomly deployed within the transmission range, indicated by $W$. In other words, a local random region around the considered node is assumed. Therefore, the probability of finding $k$ nodes in a subset $\Omega \subset W$ is given by

$$\mathbb{P}[k \text{ nodes in } \Omega] = \frac{n!}{k!(n-k)!}\left(\frac{\mu(\Omega)}{\mu(W)}\right)^k\left(1-\frac{\mu(\Omega)}{\mu(W)}\right)^{n-k} \tag{1}$$

where $\mathbb{P}$ denotes probability, $n$ is the neighborhood size in $W$, and $\mu(\cdot)$ is the standard Lebesgue measure. If we let $\Omega$ be a $d$-dimensional ball of radius $r$ originating at the considered node, $\mu(\Omega)$ is the volume of $\Omega$ and is given by $\mu(\Omega) = c_d r^d$, where

$$c_d = \frac{\pi^{\frac{d}{2}}}{\Gamma(1+d/2)} \tag{2}$$

with $\Gamma(\cdot)$ the gamma function. When $d = 1, 2$ or $3$, $c_d = 2, \pi$ and $(4/3)\pi$, respectively. For example, wireless vehicular networks can be modeled in 1-D space, a flat-Earth model requires $d = 2$, and wireless unmanned aerial vehicle communications requires $d = 3$. In this paper, all formulas are generalized in a $d$-dimensional manner for the sake of theoretical consistency.

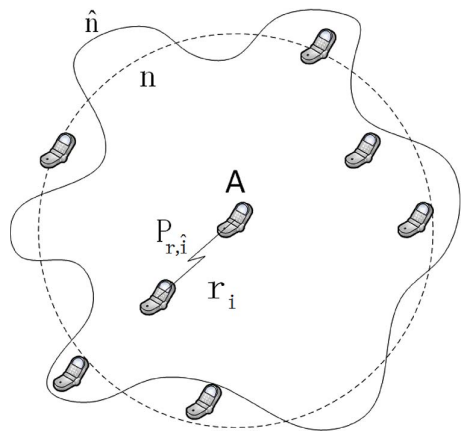

Fig. 1. Impact of the shadowing effect on node $A$. $\hat{n}$ is the estimate of the theoretical neighborhood size $n$ by counting the reachable neighbors, i.e., $\hat{n} = n + \Delta n$. By ranking the received powers at $A$, the corresponding ranking numbers $\hat{i}$ are the estimate of the ranking numbers $i$ of the ranges, where $\hat{i} = i + \Delta i$.

### B. Channel Model

The attenuation of the channel can be modeled as comprised of large-scale fading, the shadowing effect, and small-scale fading. Large-scale fading indicates that the empirical deterministic reduction in power density of an electromagnetic wave is exponentially associated with the distance when it propagates through space. We assume that the transmitted power $P_t$ is reduced through the propagation channel over a distance $r$, such that the RSS $P_r$ is given by

$$P_r = C_1 P_t \left(\frac{r_0}{r}\right)^\gamma \tag{3}$$

where $r_0 \ll r$ is the reference distance related to far-field, and $C_1$ is a nondistance-related constant that depends on the carrier frequency, the antenna gain, and the speed of light. $P_r$ and $P_t$ are both expressed in *watts*.

Depending on the environment, the PLE $\gamma$ ranges from 2 to 6 [9]. Obstacles, such as trees, buildings, and so forth, cause the actual attenuation of the received power to follow a lognormal distribution, which is also called the shadowing effect. As such, (3) has to be changed into

$$\Delta P = 10\gamma \log_{10}(r) - 10\log_{10}(C_1) - 10\gamma\log_{10}(r_0) + \chi \tag{4}$$

where $\Delta P = 10\log_{10}(P_t/P_r)$ in decibels indicates the attenuation of the signal strength, and $\chi$ follows a zero-mean *Gaussian distribution* with standard deviation $2 < \sigma < 12$. To serve the following derivations, two severe consequences of the shadowing effect should be mentioned.

1) The theoretical neighborhood size $n$ is different from the actual neighborhood size $\hat{n} = n + \Delta n$. As shown in Fig. 1 for $d = 2$, the dashed regular circle is the theoretical transmission range of node $A$. In fact, packets can be successfully received under the condition that $P_r > P_{thres}$, where $P_{thres}$ is the receiver's sensitivity. Due to the shadowing effect, the actual transmission range is irregular, as indicated by the solid line.

2) Another consequence caused by the shadowing effect is that after ranking all the received powers at node $A$, the node with the $\hat{i}$th strongest the received power $P_{r,\hat{i}}$ corresponds to the $i$th nearest neighbor at distance $r_i$, where $\hat{i} = i + \Delta i$.

When signals are being transmitted, scatterers and reflectors create several reflected paths that reach the receiver, in addition to the line of sight. This is called small-scale fading, which is nondistance related. The instantaneous received signal envelope follows the Nakagami-$m$ distribution [10], and the distribution of the instantaneous received power $p$ is, hence, given by

$$\mathbb{P}(p) = \frac{\left(\frac{m}{E(p)}\right)^m p^{m-1} e^{-\frac{mp}{E(p)}}}{\Gamma(m)} \quad (5)$$

where $m$ is the fading parameter, and a small value of $m$ indicates more fading. The measured received power $P_r$ can be obtained by taking the average over $K$ consecutive time slots of the instantaneous received power $p_k$, i.e., $P_r = (1/K)\sum_{k=1}^{K} p_k$, and thus, $\text{Var}(P_r) = [E(p_k)]^2/Km$. When $K$ is large enough, the impact of small-scale fading can be greatly eliminated. Additionally, a well-designed receiver is able to suppress the multipath effect to a great degree by using special antenna designs such as a choke ring antenna, a right-hand circularly polarized antenna, etc. Therefore, the power attenuation model in this paper is mostly subject to large-scale fading and shadowing, and hence, we will rely on (4) in the rest of this paper.

### C. Problem Statement

We are now aiming at developing a new self-estimator of the PLE. The desired properties of the proposed estimator can be summarized as *simple*, *pervasive*, *local*, *sole*, *collective*, and *secure*. *Simple* indicates that the proposed estimator should be easy to implement and carry out. *Pervasive* signals that it can be incorporated into any kind of network regardless of its design. Therefore, the only freedom left for us is to utilize the RSS. Some kind of networks might not have any external auxiliary system or access to external information, and their mutual nodal cooperations might be severely constrained. Moreover, even if there are no such constraints, adversaries can easily tamper with or forge the exchanged critical information. This requires that the estimator has to run *solely* on a single node by *collecting* the *locally* received signal strengths. By this means, a PLE can be *securely* and *locally* estimated.

As is shown in (3), the PLE $\gamma$ is strictly subject to the power attenuation and the transmitter–receiver distance. Therefore, conventional estimators in wireless localization try to obtain the PLE by introducing anchor nodes to fix the transmitter–receiver distance and by observing power attenuations. However, the desired properties of the proposed estimator determine that it is not possible to fix or to know exact transmitter–receiver distances of some of the collected RSSs. As such, we can define the problem as "How can we estimate the PLE $\gamma$ without knowing transmitter–receiver distances, i.e., merely from the local RSSs?"

## III. LINEAR REGRESSION MODEL FOR THE PATH-LOSS EXPONENT

To solve the previously mentioned problem, we introduce some new parameters. After estimating those parameters, a new linear regression model for the PLE is presented.

### A. Ranking RSSs

Let us focus on a single node and denote $P_{r,\hat{i}}$ as the $\hat{i}$th strongest power received at the considered node, where $\hat{i} = 1, 2, \ldots, \hat{n}$, i.e., $P_{r,1} \geq P_{r,2} \geq \cdots \geq P_{r,\hat{n}}$ and $r_i$ as the $i$th closest range to the considered node, where $i = 1, 2, \ldots, n$, i.e., $r_1 \leq r_2 \leq \cdots \leq r_n$. As we mentioned earlier, $\hat{i} = i + \Delta i$ is considered as an estimate of $i$, where $\Delta i$ is called the mismatch.

From (4), we can then write

$$\Delta P_{\hat{i}} = 10\gamma \log_{10}(r_i) - C_2 + \chi_i \quad (6)$$

where $\chi_i \sim \mathcal{N}(0, \sigma^2)$, $\Delta P_{\hat{i}} = 10\log_{10}(P_t/P_{r,\hat{i}})$, and $C_2 = 10\log_{10}(C_1) + 10\gamma \log_{10}(r_0)$ is a constant. We assume that all neighboring nodes transmit signals with the same power $P_t$ such that the ordered values of $P_{r,\hat{i}}$ lead to the ordered values of $\Delta P_{\hat{i}}$, i.e., we can assume that $\Delta P_1 \leq \Delta P_2 \leq \cdots \leq \Delta P_{\hat{n}}$. Admittedly, in a more realistic situation, the transmit power $P_t$ at each neighboring node might be different. However, our proposed estimators can still remain feasible in such a case, and we will come back to this issue in Section IV-D.

### B. Linear Regression Model for the PLE

From (6), we notice that $\Delta P_{\hat{i}}$ is a function of $P_t$ and $C_2$, which are both unknown. However, these can be canceled by subtracting $\Delta P_{\hat{j}}$ from $\Delta P_{\hat{i}}$, leading to $\Delta P_{\hat{i},\hat{j}} = \Delta P_{\hat{i}} - \Delta P_{\hat{j}} = 10\log_{10}(P_{r,\hat{j}}/P_{r,\hat{i}})$, which can further be written as

$$\Delta P_{\hat{i},\hat{j}} = 10\gamma \log_{10}(r_i) - 10\gamma \log_{10}(r_j) + \chi_{i,j}$$
$$= 10\gamma \log_{10}\left(\frac{r_i}{r_j}\right) + \chi_{i,j} \quad (7)$$

where $\chi_{i,j} \sim \mathcal{N}(0, 2\sigma^2)$.

Now, we define $L_i = 10\log_{10}(r_i)$ as a logarithmic function of $r_i$, and hence, $L_{i,j} = L_i - L_j = 10\log_{10}(r_i/r_j)$. Thus, (7) becomes

$$\Delta P_{\hat{i},\hat{j}} = \gamma L_{i,j} + \chi_{i,j}. \quad (8)$$

It is already apparent that if $L_{i,j}$ can be estimated, a linear regression model for the PLE can be constructed from (8). Let us denote $\widehat{L}_{\hat{i},\hat{j}}$ as the estimate of $L_{i,j}$ and $\varepsilon_{\hat{i},\hat{j}}$ as the corresponding estimation error. The linear regression model is then given by

$$\Delta P_{\hat{i},\hat{j}} = \gamma(\widehat{L}_{\hat{i},\hat{j}} - \varepsilon_{\hat{i},\hat{j}}) + \chi_{i,j}. \quad (9)$$

### C. Estimation of $L_{i,j}$

As discussed in the problem statement, it is not possible to directly obtain the transmitter–receiver distances if the estimating node solely and locally collects the RSSs. Therefore, the idea of ranking the RSSs is crucial for our method.

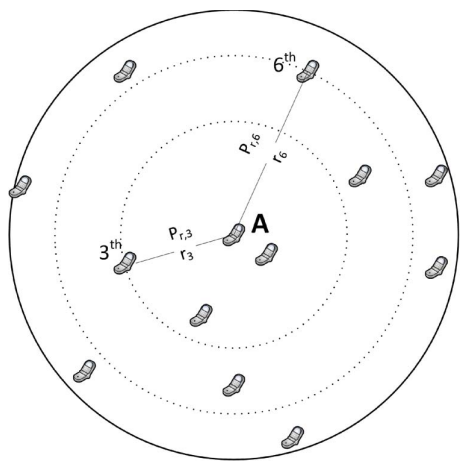

Fig. 2. In 2-D space when the shadowing effect does not impact the ranking, i.e., $\hat{i} = i$, the solid circle shows the transmit range of random node $A$, where $A$ receives 12 signal strengths from its neighbors. Its third and sixth closest neighbors lie on the dotted circles, which have $r_3$ and $r_6$ as radii, respectively. Therefore, $r_3$ has three nodes inside, whereas $r_6$ has six nodes inside. $P_{r,3}$ and $P_{r,6}$ are, respectively, the third and sixth strongest received power. $\Delta P_{3,6} = 10 \log_{10}(P_{r,6}/P_{r,3})$, and $\widehat{L}_{3,6} = (10/2)\log_{10}(3/6) \approx -1.505$. Likewise, other pairs of $\Delta P_{\hat{i},\hat{j}}$ and $\widehat{L}_{\hat{i},\hat{j}}$ can be obtained.

By ranking the values of $P_{r,\hat{i}}$, we obtain the ranking number $\hat{i}$, which will be further used to estimate the ranking numbers $i$ of the ranges, where we recall that $\hat{i} = i + \Delta i$. Additionally, it is obvious that $i$ indicates the number of nodes within the ball of radius $r_i$, which can be further exemplified in Fig. 2. Therefore, the essence of the proposed method is to use the rank numbers of $\hat{i}$ as new measurements to estimate the values of $L_{i,j}$.

Note that $L_{i,j}$ is a linear combination of $L_i$ and $L_j$. We focus on estimating $L_i$, and the estimate of $L_j$ can be obtained likewise.

Considering (1) and (2), the probability mass function of finding $i$ nodes within the $d$-ball of radius $r_i$, which is parameterized by $L_i = 10 \log_{10}(r_i)$, can be written as

$$\mathbb{P}[i|L_i] = \frac{n!}{i!(n-i)!} \left(\frac{c_d 10^{\frac{dL_i}{10}}}{\mu(W)}\right)^i \left(1 - \frac{c_d 10^{\frac{dL_i}{10}}}{\mu(W)}\right)^{n-i}. \quad (10)$$

Based on (10), to find the maximum likelihood (ML) estimator $\widehat{L}_i$, we need to force the derivative of our likelihood function to zero by

$$\frac{\partial \ln(\mathbb{P}[i|L_i])}{\partial L_i} = 0. \quad (11)$$

Therefore, by solving (11), the ML estimator $\widehat{L}_i$ can be easily obtained as

$$\widehat{L}_i = \frac{10}{d} \log_{10}\left(\frac{i\mu(W)}{nc_d}\right). \quad (12)$$

Likewise, $\widehat{L}_j$ can be obtained, and the estimate of $L_{i,j}$ is, hence, given by

$$\widehat{L}_{i,j} = \frac{10}{d} \log_{10}\left(\frac{i}{j}\right) = L_{i,j} + \varepsilon_{i,j} \quad (13)$$

where $\varepsilon_{i,j}$ is the estimation error of $\widehat{L}_{i,j}$. Plugging $\hat{i} = i + \Delta i$ and $\hat{j} = j + \Delta j$ into (13), we have

$$\widehat{L}_{\hat{i},\hat{j}} = \frac{10}{d} \log_{10}\left(\frac{\hat{i}}{\hat{j}}\right) = L_{i,j} + \varepsilon_{\hat{i},\hat{j}} \quad (14)$$

$$\varepsilon_{\hat{i},\hat{j}} = \varepsilon_{i,j} + \Delta\varepsilon_{i,j} \quad (15)$$

where $\Delta\varepsilon_{i,j} = \widehat{L}_{\hat{i},\hat{j}} - \widehat{L}_{i,j} = (10/d) \log_{10}(((i + \Delta i)/i)(j/(j + \Delta j)))$.

From (13) and (14), we even notice that $\mu(W)$, $n$, and $c_d$ disappear after subtraction. This makes the proposed estimators only subject to the RSSs and the rank numbers in $d$-dimensional space.

## IV. PATH-LOSS EXPONENT ESTIMATION

To solve the linear regression model, the total least squares (TLS) method helps us obtain the estimate of the PLE $\gamma$. However, the general solution to the TLS method turns out to be time consuming. Therefore, a closed-form solution is provided, saving computational time tremendously. Moreover, a closed-form weighted TLS method is further proposed to suppress the estimation errors, yielding a better performance.

### A. TLS Solution

As for the example in Fig. 2, node $A$ computes $\Delta P_{\hat{i},\hat{j}}$ and estimates $\widehat{L}_{\hat{i},\hat{j}}$ for all pairs of nodes within its range, i.e., $\hat{i}, \hat{j} = 1, 2, 3, \ldots, \hat{n}$. However, from (9), we notice that the RSSs are corrupted by shadowing, and the values of $\widehat{L}_{\hat{i},\hat{j}}$ are measured with errors. Therefore, the TLS method is utilized to obtain our estimate, i.e., $\widehat{\gamma}_{tls}$ [11].

We assume that the considered node has $\hat{n}$ neighbors, and all RSSs from its neighbors are ranked. Thus, we have a sample set of $\Delta P_{\hat{i},\hat{j}}$ values whose size is $N = \binom{\hat{n}}{2}$ in total. We vectorize the collected samples of $\Delta P_{\hat{i},\hat{j}}$ and the corresponding values of $\widehat{L}_{\hat{i},\hat{j}}$, which are, respectively, represented by the $N \times 1$ vectors $\mathbf{\Delta P}$ and $\widehat{\mathbf{L}}$. Then, (9) can be rewritten as

$$\mathbf{\Delta P} = \gamma(\widehat{\mathbf{L}} - \mathbf{E}) + \mathbf{X} \quad (16)$$

where $\mathbf{E}$ and $\mathbf{X}$ are, respectively, the $N \times 1$ vectors obtained by stacking the estimation errors $\varepsilon_{\hat{i},\hat{j}}$ on $\widehat{L}_{\hat{i},\hat{j}}$ and the shadowing parameters $\chi_{i,j}$. The basic idea of the TLS method is to find an optimally corrected system of equations $\mathbf{\Delta P}_{tls} = \gamma \widehat{\mathbf{L}}_{tls}$ with $\mathbf{\Delta P}_{tls} := \mathbf{\Delta P} - \mathbf{X}_{tls}$, $\widehat{\mathbf{L}}_{tls} := \widehat{\mathbf{L}} - \mathbf{E}_{tls}$, where $\mathbf{X}_{tls}$ and $\mathbf{E}_{tls}$ are, respectively, optimal perturbation vectors. Therefore, the PLE estimate $\widehat{\gamma}_{tls}$ for $\gamma$ is the solution to the optimization problem

$$\{\widehat{\gamma}_{tls}, \mathbf{X}_{tls}, \mathbf{E}_{tls}\} := \arg\min_{\gamma,\mathbf{X},\mathbf{E}} \|[\mathbf{X}\ \mathbf{E}]\|_F^2 \quad (17)$$

subject to (16), where $\|\cdot\|_F$ is the Frobenius norm.

By changing (16) into

$$\left[(\widehat{\mathbf{L}} - \mathbf{E})\ (\mathbf{\Delta P} - \mathbf{X})\right] \begin{bmatrix} \gamma \\ -1 \end{bmatrix} = 0 \quad (18)$$

we see that this is a typical low-rank approximation problem where the rank of the augmented matrix $[\widehat{\mathbf{L}}\ \mathbf{\Delta P}]$ should be optimally reduced to 1.

Therefore, we compute the singular value decomposition (SVD) of $[\widehat{\mathbf{L}}\ \Delta\mathbf{P}]$, resulting in

$$[\widehat{\mathbf{L}}\ \Delta\mathbf{P}] = \mathbf{U}\boldsymbol{\Sigma}\mathbf{V}^T$$

where $\mathbf{V}$ can be explicitly expressed as

$$\mathbf{V} = \begin{bmatrix} V_{11} & V_{12} \\ V_{21} & V_{22} \end{bmatrix}.$$

Based on the Eckart–Young theorem [12], the estimated PLE is then given by

$$\widehat{\gamma}_{tls} = -\frac{1}{V_{22}}V_{21}. \tag{19}$$

### B. Closed-Form TLS Estimation

The SVD-based method discussed in the previous section provides a general solution to the TLS problem. However, considering the complexity brought by the SVD when processing a tremendous number of samples, a simplified method is required.

Noting the linearity of (16) and the fact that the TLS method minimizes the orthogonal residuals, we can reformulate the TLS cost function as

$$J_{tls} = \frac{\|\Delta\mathbf{P} - \gamma\widehat{\mathbf{L}}\|^2}{1 + \gamma^2}. \tag{20}$$

By solving

$$\frac{\partial J_{tls}}{\partial \gamma} = \frac{\gamma^2 \widehat{\mathbf{L}}^T \Delta\mathbf{P} + \gamma(\widehat{\mathbf{L}}^T\widehat{\mathbf{L}} - \Delta\mathbf{P}^T\Delta\mathbf{P}) - \widehat{\mathbf{L}}^T\Delta\mathbf{P}}{(1+\gamma^2)^2} = 0 \tag{21}$$

we obtain two solutions, which are, respectively, given by

$$\widehat{\gamma}_1 = \eta + \sqrt{1 + \eta^2} > 0 \tag{22}$$

$$\widehat{\gamma}_2 = \eta - \sqrt{1 + \eta^2} < 0 \tag{23}$$

where $\eta = (\Delta\mathbf{P}^T\Delta\mathbf{P} - \widehat{\mathbf{L}}^T\widehat{\mathbf{L}})/2\widehat{\mathbf{L}}^T\Delta\mathbf{P}$.

In fact, optimizing (20) can also be viewed as finding a linear curve with slope $\gamma$ through the origin, in which the values of $P_{\hat{i},\hat{j}}$ and the values of $\widehat{L}_{\hat{i},\hat{j}}$ are, respectively, on the $y$-axis and the $x$-axis. See [13] for some other TLS solutions to different modified linear regression models. Therefore, it is evident that two perpendicular curves are obtained, i.e., $\widehat{\gamma}_1\widehat{\gamma}_2 = -1$. One of the solutions minimizes $J_{tls}$, whereas the other maximizes it. Considering that $\widehat{\gamma}_{tls} > 0$, the TLS-PLE estimate is obviously given by $\widehat{\gamma}_{tls} = \widehat{\gamma}_1$.

As far as computational complexity is concerned, the SVD procedure on $[\widehat{\mathbf{L}}\ \Delta\mathbf{P}]$ requires a complexity of approximately $8N^2$ to obtain $\mathbf{U}$, $\boldsymbol{\Sigma}$, and $\mathbf{V}$ [14]. If only $\mathbf{V}$ is required to estimate the PLE, the SVD-based method still has a complexity of approximately $16N$. However, our closed-form solution has only a complexity of approximately $3N$.

Compared with the SVD-based solution, we also measure the average computational time when the transmission range is 200 m. The methods are implemented in MATLAB 2012b on a Lenovo IdeaPad Y570 Laptop (Processor 2.0 GHz Intel Core i7, Memory 8 GB). As shown in Fig. 3, the computational time of the closed-form solution is greatly reduced particularly when the sample size is increased.

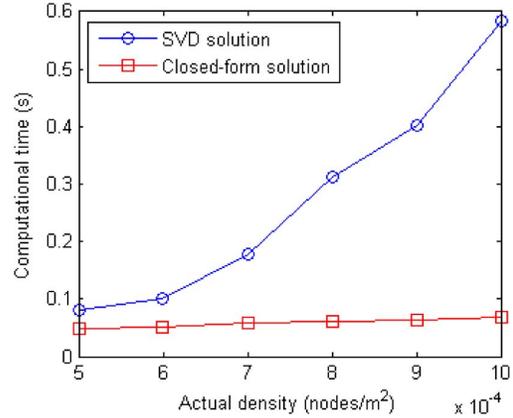

Fig. 3. Computational time of the traditional solution and the closed-form solution.

### C. Closed-Form Weighted TLS Estimation

From the aforementioned analyses, we can conclude that there are three kinds of errors impacting the PLE estimate.

1) The estimation error $\varepsilon_{i,j}$ on $\widehat{L}_{i,j}$ is subject to the spatial dynamics of the node deployment. Therefore, when increasing the actual density, such errors will be decreased.
2) The shadowing effect introduces a Gaussian error $\chi_{i,j}$, which will decrease when the sample size is increased.
3) The last kind of error is $\Delta\varepsilon_{i,j}$, which represents the mismatch between the ranking numbers of the received power and the ranges. This kind of error is subject not only to shadowing but also to the spatial dynamics of the nodes. When the actual density is increased and the nodes get closer to each other, the differences of the received power become relatively small, which leads to a large impact of shadowing on the ranking.

We propose a weighted TLS method targeting the suppression of $\Delta\varepsilon_{i,j}$. Plugging $\hat{i} = i + \Delta i$ and $\hat{j} = j + \Delta j$ into $\Delta\varepsilon_{i,j}$, we have

$$\Delta\varepsilon_{i,j} = \frac{10}{d}\log_{10}\left(\frac{\hat{i}}{\hat{i}-\Delta i}\right) - \frac{10}{d}\log_{10}\left(\frac{\hat{j}}{\hat{j}-\Delta j}\right). \tag{24}$$

By using some bounds of the natural logarithm, i.e.,

$$1 - \frac{\hat{i}-\Delta i}{\hat{i}} \leq \ln\left(\frac{\hat{i}}{\hat{i}-\Delta i}\right) \leq \frac{\hat{i}}{\hat{i}-\Delta i} - 1 \tag{25}$$

where equality is obtained when $\Delta i = 0$, bounds for $\Delta\varepsilon_{i,j}$ can be computed as

$$\frac{10\ln(10)}{d}\left(2 - \frac{\hat{i}-\Delta i}{\hat{i}} - \frac{\hat{j}}{\hat{j}-\Delta j}\right) \leq \Delta\varepsilon_{i,j}$$
$$\leq \frac{10\ln(10)}{d}\left(\frac{\hat{i}}{\hat{i}-\Delta i} + \frac{\hat{j}-\Delta j}{\hat{j}} - 2\right). \tag{26}$$

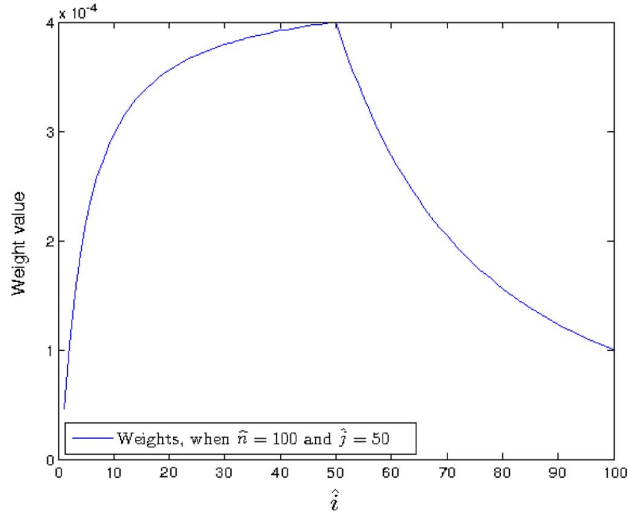

Fig. 4. TLS weights as a function of $\hat{i}$ for $\hat{n} = 100$ and $\hat{j} = 50$.

Considering that $1 \leq \hat{i} - \Delta i \leq \hat{n}$ and $1 \leq \hat{j} - \Delta j \leq \hat{n}$, we can further bound $\Delta \varepsilon_{i,j}$ as

$$\frac{10 \ln(10)}{d}\left(2 - \hat{i} - \frac{\hat{n}}{\hat{j}}\right) \leq \Delta \varepsilon_{i,j} \leq \frac{10 \ln(10)}{d}\left(\frac{\hat{n}}{\hat{i}} + \hat{j} - 2\right). \tag{27}$$

From (27), we can finally find an upper bound of $\Delta \varepsilon_{i,j}^2$ as

$$\Delta \varepsilon_{i,j}^2 \leq \frac{100 \ln(10)^2}{d^2} \max\left\{\left(\frac{\hat{n}}{\hat{i}} + \hat{j} - 2\right)^2, \left(\frac{\hat{n}}{\hat{j}} + \hat{i} - 2\right)^2\right\}. \tag{28}$$

The idea is now to assign a large weight to a sample with a small upper bound of the mismatch $\Delta \varepsilon_{i,j}^2$. Therefore, based on (28), the weights can be constructed by

$$\omega_{i,j} = \frac{1}{\max\left\{\left(\frac{\hat{n}}{\hat{i}} + \hat{j} - 2\right)^2, \left(\frac{\hat{n}}{\hat{j}} + \hat{i} - 2\right)^2\right\}}. \tag{29}$$

We plot the weights when $\hat{j} = 50$ and $\hat{n} = 100$ in Fig. 4.

By stacking the values of $\omega_{i,j}$ on the diagonal of a diagonal matrix in the same way we stack the values of $\Delta P_{\hat{i},\hat{j}}$ and the values of $\widehat{L}_{\hat{i},\hat{j}}$, we construct the $N \times N$ weight matrix $\mathbf{W}$, and then, the weighted TLS cost function can be constructed by

$$J_{wtls} = \frac{(\mathbf{\Delta P} - \gamma \widehat{\mathbf{L}})^T \mathbf{W} (\mathbf{\Delta P} - \gamma \widehat{\mathbf{L}})}{1 + \gamma^2}. \tag{30}$$

As before, the closed-form weighted TLS-PLE (WTLS-PLE) estimate is then easily given by

$$\widehat{\gamma}_{wtls} = \eta' + \sqrt{1 + \eta'^2} \tag{31}$$

where $\eta' = (\mathbf{\Delta P}^T \mathbf{W} \mathbf{\Delta P} - \widehat{\mathbf{L}}^T \mathbf{W} \widehat{\mathbf{L}})/2\widehat{\mathbf{L}}^T \mathbf{W} \mathbf{\Delta P}$.

### D. Discussions and Future Works

Here, we discuss some remaining theoretical problems and some possible issues related to real-life environments. Meanwhile, we cast light on our future works.

*1) CRLB:* The Cramér–Rao lower bound (CRLB) is very difficult to obtain for this problem. This is due to the fact that the estimation accuracy of the PLE is subject to the spatial dynamics, the shadowing, and the rank number estimate. They are all mutually related, particularly for the ranking number estimate, which does not follow any known probability density function (pdf). That is also why we selected a bound on the error to construct the weights to suppress the mismatch of the ranking numbers.

In our future work, we are looking for one-step estimation methods that can directly utilize the RSSs without the ranking procedure. To achieve that, a pdf of the RSS in an ad hoc environment is required, which considers spatial dynamics and shadowing. Based on such a pdf, a better estimator, such as the ML estimator, and the CRLB can be introduced.

*2) Different Transmit Power Values:* Previously, we assume the same transmit power $P_t$ for all the neighboring nodes, which might not be so realistic. However, assume now that the transmit power values are different. We then have to particularly estimate the transmit power $P_{t,\hat{i}}$ from the $\hat{i}$th node to calculate the path loss $\Delta P_{\hat{i}} := 10 \log_{10}(P_{t,\hat{i}}/P_{r,\hat{i}})$ and further compute $\Delta P_{\hat{i},\hat{j}} := \Delta P_{\hat{i}} - \Delta P_{\hat{j}}$. Otherwise, if we still compute $\Delta P_{\hat{i},\hat{j}} := 10 \log_{10}(P_{r,\hat{j}}/P_{r,\hat{i}})$, our estimators will become worse yet still feasible. To see that, we first need to assume an unknown average transmit power $\bar{P}_t$ and, hence, use $10 \log_{10}(P_{t,\hat{i}}) = 10 \log_{10}(\bar{P}_t) + \Delta P_{t,\hat{i}}$, where $\Delta P_{t,\hat{i}}$ is the deviation in decibels of the transmit power from the $\hat{i}$th node. Then, (9) has to be changed into

$$\Delta P_{\hat{i},\hat{j}} = \gamma(\widehat{L}_{\hat{i},\hat{j}} - \varepsilon_{\hat{i},\hat{j}}) + \chi_{i,j} + \Delta P_{t,\hat{i},\hat{j}} \tag{32}$$

where $\bar{P}_t$ can still be canceled, and $\Delta P_{t,\hat{i},\hat{j}} := \Delta P_{t,\hat{i}} - \Delta P_{t,\hat{j}}$. Obviously, although $\mathbf{X}$ in (16) has to become the vector of $\chi_{i,j} + \Delta P_{t,\hat{i},\hat{j}}$ values, our proposed estimators can still estimate the PLE since the general form of (16) remains the same.

Hence, if we assume that $\Delta P_{t,\hat{i},\hat{j}}$ is Gaussian distributed, $\chi_{i,j} + \Delta P_{t,\hat{i},\hat{j}}$ is still a zero-mean Gaussian variable, which means that the different transmit power values can equivalently be considered as a more severe shadowing impact. Therefore, for convenience, we still assume the same transmit power in this paper.

*3) Directional PLE Estimation:* Another practical problem is that the PLE sometimes varies over different directions while we previously assume that the PLE is omnidirectionally the same. To cope with this problem, we discuss and can extend our proposed estimators with a directional PLE estimation.

As shown in Fig. 5, we assume that only the RSSs from the nodes within the angular window $\phi$ are subject to the same PLE. Hereby in (1), $W$ has to become the actual transmission range bounded by the angle $\phi$, e.g., $W_\phi$, whereas $\Omega$ becomes the corresponding sector $\Omega_\phi$ with radius $r$. The volume of $\Omega_\phi$ then becomes $\mu(\Omega_\phi) := c_{d,\phi} r^d$, where for $d = 1, 2, 3$, we have $c_{1,\phi} := 1$, $c_{2,\phi} := \phi/2$, and $c_{3,\phi} := (2\pi/3)(1 - \cos \phi)$. Since

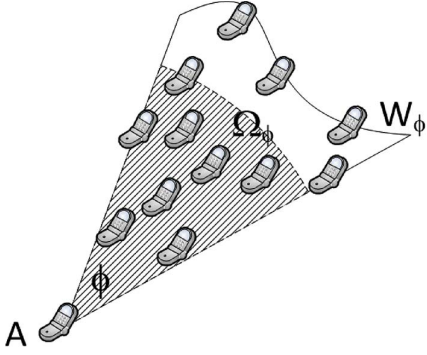

Fig. 5. Demonstration of the directional PLE estimation in $\mathbb{R}^2$. $A$ is the considered node collecting the RSSs from within the angle $\phi$. $W_\phi$ is the actual transmission range bounded by $\phi$, and the shaded area $\Omega_\phi$ is the corresponding sector with radius $r$.

the nodes are still randomly deployed within $W_\phi$, compared with (10), we can, hence, similarly write

$$\mathbb{P}[i|L_i] = \frac{n!}{i!(n-i)!} \left(\frac{c_{d,\phi} 10^{\frac{dL_i}{10}}}{\mu(W_\phi)}\right)^i \left(1 - \frac{c_{d,\phi} 10^{\frac{dL_i}{10}}}{\mu(W_\phi)}\right)^{n-i}. \quad (33)$$

Although the estimate of $L_i$ has to be changed into

$$\widehat{L}_i = \frac{10}{d} \log_{10}\left(\frac{i\mu(W_\phi)}{n c_{d,\phi}}\right) \quad (34)$$

the estimate of $L_{i,j}$, however, remains the same, i.e., $\hat{L}_{i,j} := \hat{L}_i - \hat{L}_j$, since $\mu(W_\phi)$, $n$, and $c_{d,\phi}$ will be canceled. Therefore, the rest of the theoretical derivations remain the same, and our estimators are still feasible.

To achieve a directional PLE estimate, we only have to constrain the RSS sample set within a certain angular window $\phi$, and our proposed estimators can estimate the PLE for the given direction. Of course, to achieve the same accuracy, the directional PLE estimator has to collect more samples than the omnidirectional PLE estimator. Again in this paper, for convenience, we assume the same PLE for all directions.

## V. SIMULATIONS

Here, we simulate our two proposed PLE estimators in 2-D space, and we leave real-life experiments as future work. Two simulations are conducted to study their performance, with different shadowing impacts and with different actual densities.

We also compare them with the PLE estimator based on the cardinality of the transmitting set (C-PLE) proposed in [1]. The C-PLE requires changing the receiver's sensitivity from $P_{thres1}$ to $P_{thres2}$ and evaluating the corresponding cardinalities $n_1$, $n_2$ of the transmitting set, namely, the different theoretical neighborhood sizes. Thus, considering shadowing, C-PLE is given in 2-D space by

$$\widehat{\gamma}_c = \frac{2 \ln\left(\frac{P_{thres2}}{P_{thres1}}\right)}{\ln\left(\frac{\hat{n}_1}{\hat{n}_2}\right)} \quad (35)$$

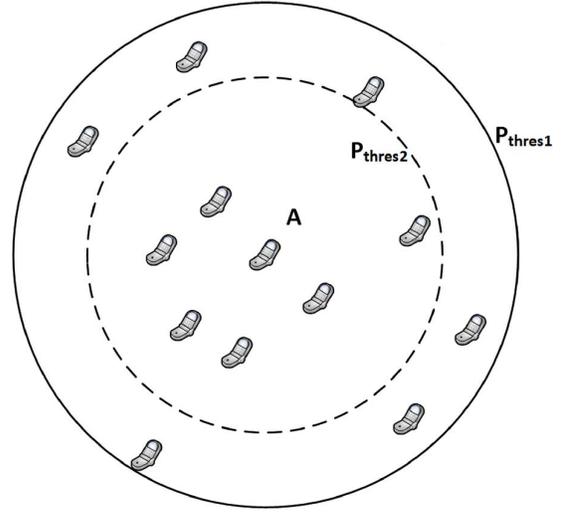

Fig. 6. Demonstration of the C-PLE estimator. Node $A$ changes its receiver's sensitivity from $P_{thres1}$ to $P_{thres2}$. The solid circle and the dashed circle are, respectively, the transmission ranges related to $P_{thres1}$ and $P_{thres2}$. The corresponding neighborhood sizes are $\hat{n}_1 = 12$ and $\hat{n}_2 = 6$ in this figure. The estimated PLE can be obtained from (35).

TABLE I
VALUES OF THE PARAMETERS USED IN THE SIMULATIONS

| Parameter | Value |
|---|---|
| Dimension | $d = 2$ |
| Carrier frequency | 2401 MHz |
| Receiver sensitivity | For TLS-PLE and WTLS-PLE, $P_{thres}$ is adjusted to have a theoretical transmission range of $200\ m$. For C-PLE, $P_{thres1} = P_{thres}$ and $P_{thres2} = 2P_{thres}$. |
| Number of trials | 100 |

where $\hat{n}_1$ and $\hat{n}_2$ are the corresponding actual neighborhood sizes. Fig. 6 gives an example of the C-PLE estimator. In our simulations, we set $P_{thres2} = 2P_{thres1}$.

To avoid any border effect, our simulations take place in a very large area, where nodes are randomly deployed. The estimated PLE is only considered for a single node somewhere in the center of the network, rather than for every node in the wireless network. The Monte Carlo method is used to generate the results by repeatedly deploying nodes. The general settings are shown in Table I.

The normalized root mean square error (RMSE) is adopted to present the accuracy of the estimator. In this paper, the normalized RMSE is defined by $\sqrt{(1/N_{trials})\sum_{i=1}^{N_{trials}}[(\widehat{\gamma}(i)-\gamma)/\gamma]^2}$, where $N_{trials}$ is the number of simulation trials, $\widehat{\gamma}(i)$ is the estimate of the PLE in the $i$th trial, and $\gamma$ is the actual PLE.

### A. Impact of Shadowing

This simulation is conducted when the actual density is set as 0.005 node/m². Three estimators are studied with an increasing standard deviation of shadowing and an increasing actual PLE. Observing Fig. 7, we can conclude the following.

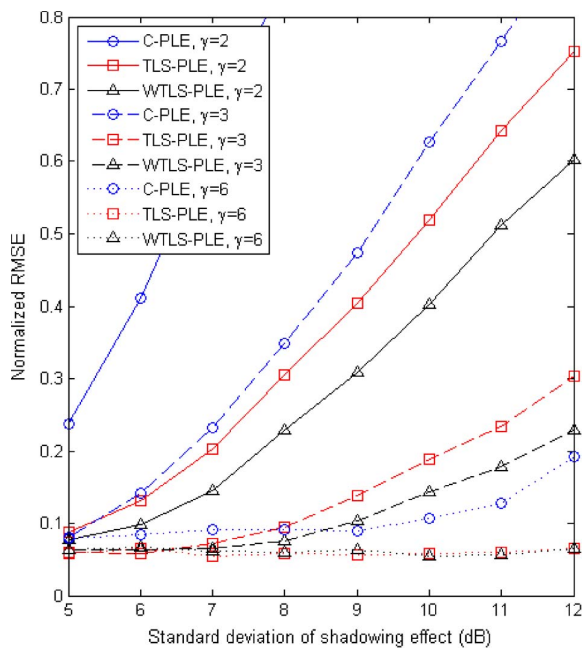

Fig. 7. Performance of different PLE estimators with an increasing standard deviation of shadowing.

1) Our two proposed methods outperform the C-PLE estimator. This can be easily understood from the fact that our methods consider received power from all neighbors rather than only using two neighborhood sizes. Moreover, the TLS procedure helps minimize the three kinds of errors mentioned earlier.
2) When the shadowing effect becomes more severe, the accuracy of the three estimators decreases. For the C-PLE, the accuracy mainly depends on the absolute deviation of the actual neighborhood size $|\Delta n| = |\hat{n} - n|$. The shadowing increases such an absolute deviation, thus leading to worse accuracy. For our methods, the shadowing impacts the accuracy by increasing $|\chi_{i,j}|$ and by disrupting the matches between the rank numbers of the received power and the ranges, i.e., by increasing $|\Delta \varepsilon_{i,j}|$.
3) Surprisingly, the performance of the estimators becomes better in a harsher environment, i.e., when the actual PLE is high. This is due to the fact that a high PLE causes relatively large differences between the received powers, which makes the shadowing effect more tolerable. It is better explained in Fig. 8. To be specific for our methods, when the PLE is small, the accuracy is subject to the three kinds of errors $\chi_{i,j}$, $\varepsilon_{i,j}$, and $\Delta \varepsilon_{i,j}$. However, when the actual PLE is increased, the matches of the rank numbers are more accurate, i.e., $|\Delta \varepsilon_{i,j}|$ decreases.
4) The WTLS-PLE has a better performance than the TLS-PLE, particularly under a small PLE. Meanwhile, the improvement of the WTLS-PLE is not so obvious compared with the TLS-PLE when the PLE is high. This is understandable from the fact that the WTLS-PLE is particularly targeted at suppressing $\Delta \varepsilon_{i,j}$, the improvement is, hence, insignificant when $\Delta \varepsilon_{i,j}$ is decreased, which has already been pointed out in the previous conclusion.

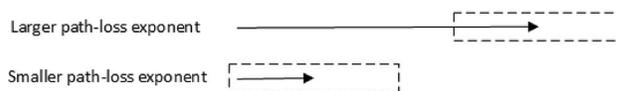

Fig. 8. Length of the arrow indicates the RSS reduction $\Delta P$, and the dashed rectangles show the shadowing effect $\chi$. Considering shadowing means that the arrows can end up anywhere within the rectangles. The width of the rectangle indicates the severity of the shadowing. Under the same transmitter–receiver distance, the arrow with a smaller PLE is shorter and thus easier to be impacted by the shadowing effect. Therefore, under a high PLE, the matching between the ranking numbers of the received power and the ranges is not so easily disrupted in the TLS-PLE and the WTLS-PLE. Likewise, the shadowing also becomes more tolerable when estimating the theoretical neighborhood size in the C-PLE.

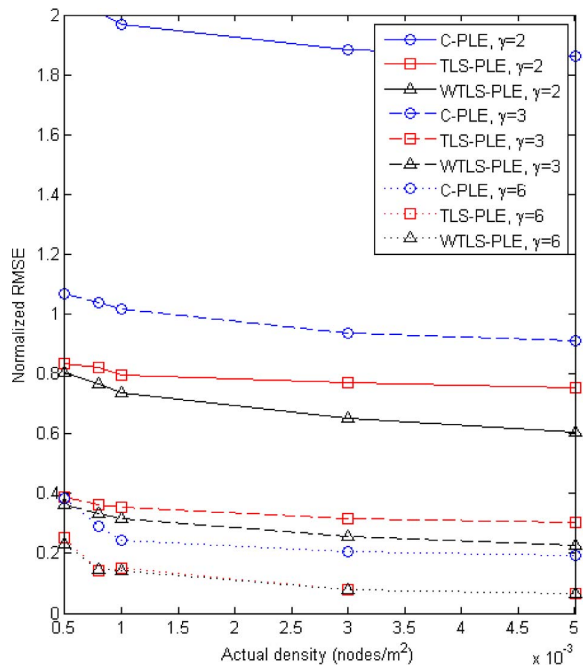

Fig. 9. Performance of three considered estimators with an increasing actual density.

### B. Impact of the Actual Density

Since the estimation error $\varepsilon_{i,j}$ of $\widehat{L}_{i,j}$ is related to the actual density, we are interested in how the actual density impacts the accuracy in this section. The transmission range is fixed at 200 m, and a 12-dB standard deviation of the shadowing is considered. In Fig. 9, we can see that, compared with the impact of shadowing, the impact of the actual density is relatively small. Additionally, when more samples are collected, the WTLS-PLE has a larger improvement on the accuracy by suppressing $\Delta \varepsilon_{i,j}$.

## VI. APPLICATIONS

The PLE plays a very significant role in many kinds of wireless networks. Due to the difficulties in locally and solely estimating the PLE, only a few techniques are able to utilize PLE measurements in their designs. However, the proposed PLE estimation approaches tackle such issues. Here, we detail some applications and discuss the significance of our PLE self-estimation schemes.

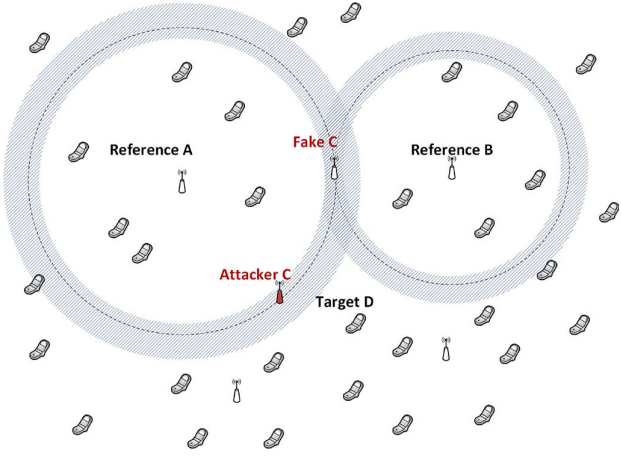

Fig. 10. *Attacker C* reports its fake location at *fake C*. Both reference nodes, such as *reference A* and *reference B*, and the target node *target D* can self-estimate the PLE. Based on the self-estimated PLE and the location information, the shaded area can be constructed as the trust region for detecting an attacker, outside of which *attacker C* will be detected.

### A. Secure RSS-Based Localization

Due to our PLE self-estimation schemes, either the reference node or the target node can solely and independently estimate the PLE. Therefore, an adversary cannot launch an attack on the PLE estimation by spoofing. For instance, as shown in Fig. 10, even if there is a cheating reference node maliciously reporting its fake location, e.g., *attacker C* registering itself at *fake C*, the PLE can still be accurately estimated. Apart from making the RSS-based localization more robust to the spoofing attack, this also enables every node to detect and locate the cheating reference node.

*1) Strategy for Detecting Cheating Reference Nodes:* To explicitly illustrate the strategy, we first explain each one's role, and the detection algorithm will be described afterward.

- Each *reference node* knows its own location and is skeptical about any reported location from the other reference nodes.
  — It periodically broadcasts its own location and self-estimates the PLE simultaneously.
  — It keeps listening to the messages broadcasted by the other reference nodes, reading the RSSs and their corresponding reported locations.
  — It detects the attackers according to the self-estimated PLE, the RSSs, the reported locations, and its own location. The detection algorithm will be discussed later. As soon as an attacker is detected, it will announce the detection as well as the corresponding RSS from the attacker by broadcasting.
  — In case some cheating reference nodes spoof the attacker announcement, an announced attacker needs to be further confirmed as a true attacker. To be confirmed as a true attacker, the announced attacker has to be announced more than $T$ times, where $T$ depends on the total number of reference nodes and the detection sensitivity. When the announced attacker is confirmed as a true attacker, the corresponding announced RSSs from the attacker at at least $d+1$ different reference nodes can further be used to locate the attacker.
- Each *target node* only listens and is invisible to the other nodes.
  — It keeps listening to all information broadcasted by the reference nodes. In the meantime, the PLE is self-estimated.
  — It discovers the true attackers from the message broadcasted by the reference nodes and discards the RSSs from the true attackers.
  — Then, it can accurately and safely locate itself with the rest of the RSSs.

*2) Algorithm for Detecting Cheating Reference Nodes:* To complete the strategy, the algorithm for detecting the cheating reference nodes is essential. For an explicit demonstration, an example is shown in Fig. 10. Let us denote the locations of *reference A*, *reference B*, *attacker C*, *fake C*, and *target D*, respectively, as $\mathbf{s}_A$, $\mathbf{s}_B$, $\mathbf{s}_C$, $\mathbf{s}_{C'}$, and $\mathbf{s}_D$. To detect *attacker C*, we need to test two hypotheses, which are respectively defined as

$$\mathcal{H}_0 : \mathbf{s}_C \text{ and } \mathbf{s}_{C'} \text{ are the same location} \quad (36)$$
$$\mathcal{H}_1 : \mathbf{s}_C \text{ and } \mathbf{s}_{C'} \text{ are different locations.} \quad (37)$$

The detection algorithm can be carried out with the following procedure.

a) First, a reference RSS from the suspected reference node needs to be calculated based on the self-estimated PLE, the reported location, and the location itself of the detecting node. For example, recalling the definition of RSS, the reference RSS at *reference B* from *attacker C* can be calculated in decibels as

$$P'_{r,C'B} = C_3 - 10\hat{\gamma}_B \log_{10}\left(\|\mathbf{s}_{C'} - \mathbf{s}_B\|\right) \quad (38)$$

where

$$C_3 = 10\log_{10}(P_t) + 10\log_{10}(C_1) + 10\hat{\gamma}_B \log_{10}(r_0)$$

and $\hat{\gamma}_B$ is the self-estimated PLE at $\mathbf{s}_B$.

b) Second, the actual RSSs from the suspected reference node are recorded over time to construct our observation set by subtracting the reference RSS. For example, *reference B* records the observation at time $i$, which is given by

$$\Delta P^i_{r,CB} = P^i_{r,CB} - P'_{r,C'B} \quad (39)$$

where $P^i_{r,CB}$ is the actual RSS in decibels at time $i$ from *attacker C* and $\Delta P^i_{r,CB} \sim \mathcal{N}(\mu_B, \sigma^2)$. If *attacker C* and *fake C* have the same range, then $\mu_B = 0$; otherwise, $\mu_B \neq 0$.

Since only the range can be tested, we need two different hypotheses for range testing, which are given by

$$\mathcal{H}'_0 : \mu_B = 0 \tag{40}$$
$$\mathcal{H}'_1 : \mu_B \neq 0. \tag{41}$$

Considering the fact that *attacker C* and *fake C* might also have the same range to a reference node, e.g., to *reference A* in Fig. 10, we hence have the relations $\mathcal{H}_0 \subset \mathcal{H}'_0$ and $\mathcal{H}'_1 \subset \mathcal{H}_1$. This means that if $\mathcal{H}'_1$ is tested, *attacker C* is certainly detected, whereas if $\mathcal{H}'_0$ is tested, we might fail to detect the attacker. However, we now focus on testing $\mathcal{H}'_1$, and the detection failure in $\mathcal{H}'_0$ will be discussed later.

c) Finally, by using the Neyman–Pearson lemma [15], $\mathcal{H}'_1$ can be tested from the average observation over $I$ time slots. For example, the observation at *reference B* is given by $\rho = (\sum_{i=1}^{I} \Delta P^i_{r,CB})/I$. If we wish to test at 95% accuracy, the critical region for the observation is given by

$$C = \left\{ (P^1_{r,CB}, P^2_{r,CB}, \ldots, P^I_{r,CB}) : \right.$$
$$\left. \rho \leq -1.96\sigma/\sqrt{I}, \rho \geq 1.96\sigma/\sqrt{I} \right\}. \tag{42}$$

Equivalently, we can also use the critical region, i.e.,

$$C = \left\{ (P^1_{r,CB}, P^2_{r,CB}, \ldots, P^I_{r,CB}) : \rho^2 \geq 3.84\sigma^2/I \right\} \tag{43}$$

which considers the Chi-squared distribution with $\rho^2$ as observation.

3) *Discussions:*

a) The shadowing deviation $\sigma$ is required for the Neyman–Pearson test, which can be obtained by empirical training.
b) The detection failure in $\mathcal{H}'_0$ can easily be noticed when reference nodes work in a cooperative fashion according to the detection strategy. Since every reference node detects and announces the attackers, such a detection failure can be somehow corrected by listening to the announced information flooding in the network. Therefore, the detection algorithm can be improved by introducing a new cooperative algorithm. For example, according to the observations from multiple nodes, an attacker can still be detected even if such a detection failure in $\mathcal{H}'_0$ occurs.
c) Considering shadowing, the complement of the critical region corresponds to a trust region of the detecting node in space, in which the detected node will be trusted. As shown in Fig. 10, two shaded areas, respectively, indicate the trust regions of *reference A* and *reference B*. *Attacker C* resides outside the trust region of *reference B* but inside that of *reference A*. Therefore, *attacker C* will be detected by *reference B* but not by *reference A*. The size of the trust region depends on the severity of the shadowing.
d) The cheating node can also jeopardize this system by maliciously announcing a credible reference node as an attacker. In most cases, the credible reference nodes outnumber the attackers. Hence, the attackers can still be smartly distinguished. However, if the attackers have the majority, a more robust strategy might be required.

### B. Energy-Efficient Routing

Since the path loss over a channel exponentially increases with the distance, multihop communications becomes a better option than single-hop communications to prolong the network lifetime. Routing is hence aimed at finding an efficient path to the destination to minimize the power consumption. It is well known that a routing path is better to be chosen through an area where the PLE is small. However, alternatively, here, we consider the $k$th nearest neighbor routing protocol to illustrate the significance of the PLE.

From (1), if considering the local random region $W$ around the considered node $A$ as a $d$-dimensional ball of radius $R$, i.e., $\mu(W) = c_d R^d$, the distribution of the distance $r_k$ to the $k$th nearest neighbor is given by [16]

$$\mathbb{P}(r_k|k) = \frac{d}{r_k B(n-k+1,k)} \left(\frac{r_k^d}{R^d}\right)^k \left(1 - \frac{r_k^d}{R^d}\right)^{n-k} \tag{44}$$

where $B(x,y) = \int_0^1 t^{x-1}(1-t)^{y-1}dt = \Gamma(x)\Gamma(y)/\Gamma(x+y)$ is the beta function. To avoid the singularity issue of (3), the received power at the $k$th nearest neighbor can also be given by

$$P_{r,k} = P_{r,0} \left(\frac{r_0}{r_k}\right)^{\gamma_A} \tag{45}$$

where $P_{r,0}$ is the received power at the reference distance $r_0 < r_k \; \forall k$, and $\gamma_A$ is the PLE at the location of node $A$. Let us denote the path loss to the $k$th nearest neighbor as $\mathfrak{L}_k := P_{r,0}/P_{r,k} = r_k^{\gamma_A}/r_0^{\gamma_A}$. We commonly assume $r_0 = 1$ m, and thus, $\mathfrak{L}_k := r_k^{\gamma_A}$. From (44), we can obtain the expectation of $\mathfrak{L}_k$ for a single hop to the $k$th nearest neighbor, which can be given by

$$E(\mathfrak{L}_k) = \frac{R^{\gamma_A} B(k+\gamma_A/d, n-k+1)}{B(n-k+1,k)}$$
$$= \frac{R^{\gamma_A} \Gamma(n+1)}{\Gamma(n+\gamma_A/d+1)} \frac{\Gamma(k+\gamma_A/d)}{\Gamma(k)}. \tag{46}$$

From (46), we particularly focus on $\partial E(\mathfrak{L}_k)/\partial k$ to study the efficiency of increasing $k$, which is given by

$$\frac{\partial E(\mathfrak{L}_k)}{\partial k} = \frac{R^{\gamma_A} \Gamma(n+1)}{\Gamma(n+\frac{\gamma_A}{d}+1)} \frac{\Gamma\left(k+\frac{\gamma_A}{d}\right)\left(\psi\left(k+\frac{\gamma_A}{d}\right) - \psi(k)\right)}{\Gamma(k)} \tag{47}$$

where $\psi(x) = \Gamma'(x)/\Gamma(x)$ is the polygamma function. We denote $\alpha = \gamma_A/d$ and plot the $k$-related part of (47), i.e., $f(k) = \Gamma(k+\alpha)(\psi(k+\alpha) - \psi(k))/\Gamma(k)$ in Fig. 11. When $\alpha < 1$, $\partial E(\mathfrak{L}_k)/\partial k$ decreases with $k$, which means that it takes less extra power every time $k$ is increased. As a conclusion, a single long-hop communication link is more energy efficient, as long as $\gamma_A < d$, which is also briefly pointed out in [16].

To be more realistic, we also conduct a numerical simulation for the $k$th nearest neighbor routing, in which the shadowing effect is also considered. We introduce the average path loss for a single link, which is denoted by $\overline{\mathfrak{L}}_k = \mathfrak{L}_k/k$. Two-dimensional space is considered with a density of 0.001 nodes/m$^2$. As is shown in Fig. 12, as long as the PLE is smaller than 2, the average path loss decreases with $k$, and a single long-hop link becomes energy efficient. Additionally, in the presence of log-normal shadowing, $\overline{\mathfrak{L}}_k$ becomes larger than when there is no shadowing. Such an increase also becomes larger with a large PLE.

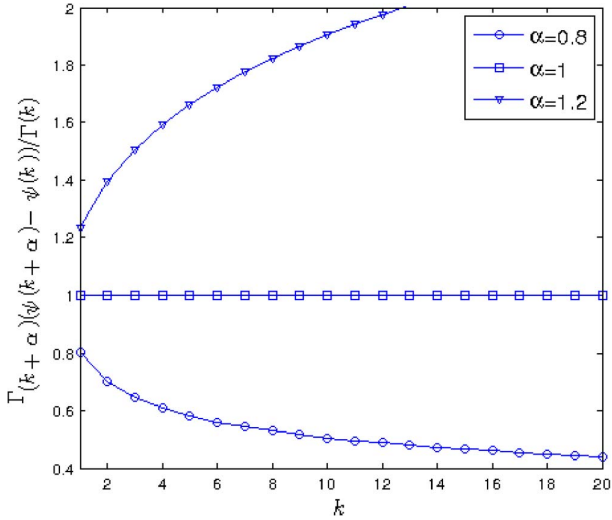

Fig. 11. Efficiency of single-hop communications: A small value indicates a smaller power cost by increasing $k$, i.e., a high power efficiency.

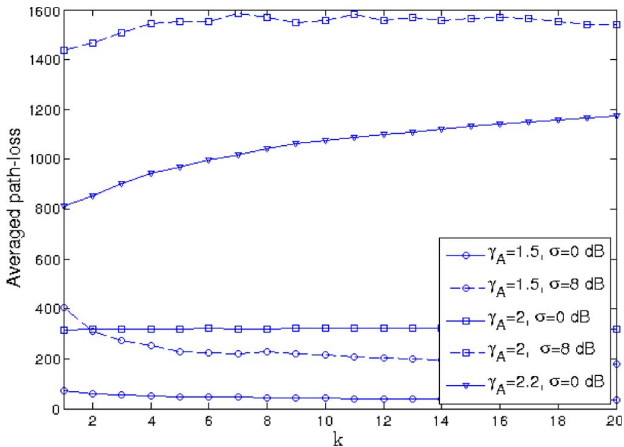

Fig. 12. Numerical results of the $k$th nearest neighbor routing in 2-D space.

Many other interesting results have been obtained. However, due to space limitations, no more tautology will be presented. It is already evident that the efficiency of the $k$th nearest neighbor routing protocol highly relies on the actual PLE. Therefore, the principles for designing such a routing protocol should involve the PLE estimation. In a nutshell, an accurate estimate of $\gamma_A$ is hence necessary for designing an efficient routing protocol.

### C. Other Applications

To further illustrate some applications of the proposed PLE estimators, we need to explicitly explain how the PLE affects the network operation. The PLE has a multidimensional effect on the performance of the whole system for wireless networking.

1) It determines the quality of the signals at the receivers and thus impacts the physical (PHY) layer. This is because the PLE controls not only the RSS but the interference the nodes create for the other receivers as well. Since the signal-to-interference-plus-noise ratio is decisive for the channel capacity and the performance of decoding, the PLE is essential for designing the PHY layer.
2) It determines the transmission range and thus impacts the network (NET) layer and the media access control (MAC) layer. The transmission range, together with the neighborhood size, which is also determined by the PLE, affects the performance of routing and the connectivity in the NET layer. When the number of nodes within the transmission range of a node increases, the contention in the MAC layer consequently becomes more severe, and thus, congestion of the network will occur. As a consequence, the ability of delivering the packet will be affected.
3) It determines the energy consumption for transmission links and thus impacts the lifetime of networking. To guarantee the efficiency of wireless networking, the transmit power should be smartly controlled to compensate for the energy loss in the transmission links. Considering that the battery is strictly limited in, e.g., wireless sensor networks, the PLE is rather significant to those protocols aiming at prolonging the network lifetime.

Based on the aforementioned reasons, some other applications can be listed.

1) *Relay nodes* are recently drawing much attention [17], and the mobile relay nodes are even more flexible and more convenient [18]. Since the PLE is one of the key criteria for energy-efficient routing, relay nodes should be deployed or move to the place where the PLE is relatively small. The relay nodes can also benefit from the low-PLE location to save the battery. Therefore, relay nodes have to be able to estimate the PLE.
2) *Energy harvesting* relies on ambient sources such as solar, wind, and kinetic activities, aiming at prolonging the network lifetime. Particularly, among those sources, radio-frequency signals can also be used to charge the battery of wireless sensors [19]. Its application is also extended to cognitive radio in [20]. The PLE directly determines the efficiency of harvesting and the size of the harvesting zone. The time slot for harvesting could be adaptive according to the PLE changes. Therefore, the PLE estimation is very significant when the surrounding communication environment is changing or the harvesting node is mobile.
3) *Power control* requires distributedly choosing an appropriate transmit power for each packet at each node. This is because of the fact that the transmit power affects the wireless networking in the same way as the PLE does [21]. Since the PLE is different at different locations, an efficient power control scheme also needs to distributedly and locally consider the PLE. Therefore, our proposed estimators can be integrated into power control to yield a better performance.

## VII. Conclusion

Two self-estimators for the PLE have been proposed in this paper, in which each node can solely and locally estimate the PLE merely by collecting the RSSs. They rely neither on external auxiliary systems nor on any information of the wireless network. Their simplicity makes them feasible for any kind of wireless network.

To better describe our estimators, a new linear regression model for the PLE has been introduced. Our closed-form TLS method can solve this linear regression model. Compared with the SVD-based solution, our estimator tremendously saves computational time. Moreover, a weighted TLS method is also designed to better suppress the estimation errors.

Simulations present the accuracy of our estimators and demonstrate that the shadowing effect dominantly influences the estimation error. By analyzing the performance of the estimators, it is interesting to observe that the estimators work better in harsh communication environments, where the PLE is high.

We have also discussed the significance of our PLE self-estimators by illustrating some potential applications and have brought the dawn to some relevant future research.

## Acknowledgment

The authors would like to sincerely thank the generous help from E. Onur, a member of IEEE and ACM (e-mail: onure@ieee.org).

## References

[1] S. Srinivasa and M. Haenggi, "Path loss exponent estimation in large wireless wetworks," in *Proc. Inf. Theory Appl. Workshop*, 2009, pp. 124–129.
[2] N. Patwari *et al.*, "Locating the nodes: Cooperative localization in wireless sensor networks," *IEEE Signal Process. Mag.*, vol. 22, no. 4, pp. 54–69, Jul. 2005.
[3] G. Mao, B. D. O. Anderson, and B. Fidan, "Path loss exponent estimation for wireless sensor network localization," *Comput. Netw.*, vol. 51, no. 10, pp. 2467–2483, Jul. 2007.
[4] N. Salman, M. Ghogho, and A. Kemp, "On the joint estimation of the RSS-based location and path-loss exponent," *IEEE Wireless Commun. Lett.*, vol. 1, no. 1, pp. 34–37, Feb. 2012.
[5] N. Salman, A. Kemp, and M. Ghogho, "Low complexity joint estimation of location and path-loss exponent," *IEEE Wireless Commun. Lett.*, vol. 1, no. 4, pp. 364–367, Aug. 2012.
[6] X. Li, "RSS-based location estimation with unknown pathloss model," *IEEE Trans. Wireless Commun.*, vol. 5, no. 12, pp. 3626–3633, Dec. 2006.
[7] M. Gholami, R. Vaghefi, and E. Strom, "RSS-based sensor localization in the presence of unknown channel parameters," *IEEE Trans. Signal Process.*, vol. 61, no. 15, pp. 3752–3759, Aug. 2013.
[8] G. Wang, H. Chen, Y. Li, and M. Jin, "On received-signal-strength based localization with unknown transmit power and path loss exponent," *IEEE Wireless Commun. Lett.*, vol. 1, no. 5, pp. 536–539, Oct. 2012.
[9] T. Rappaport, *Wireless Communications: Principles and Practice*, 2nd ed. Upper Saddle River, NJ, USA: Prentice-Hall, 2001.
[10] N. Nakagami, "The m-distribution, a general formula for intensity distribution of rapid fading," in *Statistical Methods in Radio Wave Propagation*, W. G. Hoffman, ed. Oxford, U.K.: Pergamon, 1960.
[11] I. Markovsky and S. Van Huffel, "Overview of total least-squares methods," *Signal Process.*, vol. 87, no. 10, pp. 2283–2302, Oct. 2007. [Online]. Available: http://dx.doi.org/10.1016/j.sigpro.2007.04.004
[12] C. Eckart and G. Young, "The approximation of one matrix by another of lower rank," *Psychometrika*, vol. 1, no. 3, pp. 211–218, Sep. 1936. [Online]. Available: http://dx.doi.org/10.1007/BF02288367
[13] I. Petras and D. Bednarova, "Total least squares approach to modeling: A Matlab toolbox," *Acta Montanistica Slovaca*, vol. 15, no. 2, pp. 158–178, 2010.
[14] G. H. Golub and C. F. Van Loan, *Matrix Computations*, 3rd ed. Baltimore, MD, USA: The Johns Hopkins Univ. Press, 1996.
[15] P. Hoel, S. Port, and C. Stone, *Introduction to Statistical Theory*, ser. Houghton Mifflin research series, Boston, MA, USA: Houghton-Mifflin, 1971. [Online]. Available: http://books.google.nl/books?id=6hPvAAAAMAAJ
[16] S. Srinivasa and M. Haenggi, "Distance distributions in finite uniformly random networks: Theory and applications," *IEEE Trans. Veh. Technol.*, vol. 59, no. 2, pp. 940–949, Feb. 2010.
[17] S. Misra, S. D. Hong, G. Xue, and J. Tang, "Constrained relay node placement in wireless sensor networks: Formulation and approximations," *IEEE/ACM Trans. Netw.*, vol. 18, no. 2, pp. 434–447, Apr. 2010.
[18] A. Venkateswaran, V. Sarangan, T. La Porta, and R. Acharya, "A mobility-prediction-based relay deployment framework for conserving power in MANETs," *IEEE Trans. Mobile Comput.*, vol. 8, no. 6, pp. 750–765, Jun. 2009.
[19] T. Le, K. Mayaram, and T. Fiez, "Efficient far-field radio frequency energy harvesting for passively powered sensor networks," *IEEE J. Solid-State Circuits*, vol. 43, no. 5, pp. 1287–1302, May 2008.
[20] S. Lee, R. Zhang, and K. Huang, "Opportunistic wireless energy harvesting in cognitive radio networks," *IEEE Trans. Wireless Commun.*, vol. 12, no. 9, pp. 4788–4799, Sep. 2013.
[21] V. Kawadia and P. Kumar, "Principles and protocols for power control in wireless ad hoc networks," *IEEE J. Sel. Areas Commun.*, vol. 23, no. 1, pp. 76–88, Jan. 2005.

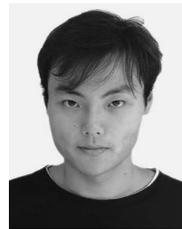

**Yongchang Hu** (M'13) was born in Xi'an, China, in 1988. He received the B.Sc. and M.Sc. degrees from Northwestern Polytechnical University, Xi'an, in 2010 and 2013, respectively. He is currently working toward the Ph.D. degree with the Circuits and Systems (CAS) Group, Department of Microelectronics, Delft University of Technology, Delft, The Netherlands.

His research interests lie in signal processing for wireless communications, modeling radio propagation channels, random network analysis, and wireless source localization.

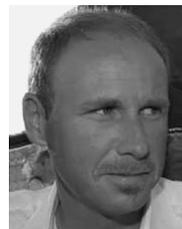

**Geert Leus** (F'12) received the electrical engineering degree and the Ph.D. degree in applied sciences from the Katholieke Universiteit Leuven, Leuven, Belgium, in 1996 and 2000, respectively.

He is currently an "Antoni van Leeuwenhoek" Full Professor with the Faculty of Electrical Engineering, Mathematics and Computer Science, Delft University of Technology, Delft, The Netherlands. His research interests are in the area of signal processing for communications.

Dr. Leus received the 2002 IEEE Signal Processing Society Young Author Best Paper Award and the 2005 IEEE Signal Processing Society Best Paper Award. He was the Chair of the IEEE Signal Processing for Communications and Networking Technical Committee and an Associate Editor of the IEEE TRANSACTIONS ON SIGNAL PROCESSING, the IEEE TRANSACTIONS ON WIRELESS COMMUNICATIONS, the IEEE SIGNAL PROCESSING LETTERS, and the EURASIP *Journal on Advances in Signal Processing*. He is currently a Member-at-Large of the Board of Governors of the IEEE Signal Processing Society and a member of the IEEE Sensor Array and Multichannel Technical Committee. He also serves as the Editor-in-Chief of the EURASIP *Journal on Advances in Signal Processing*.